\documentclass[twocolumn]{aa}
\usepackage{graphicx}
\usepackage{array}
\usepackage{amsfonts}
\usepackage{natbib}

\def\llabel#1{\label{#1}} 

\graphicspath{{./figures/}}
\def\deg{{^\circ}}

\def\sun{\odot}

\def\m@th{\mathsurround=0pt}
\def\EQM#1{\vcenter{\normalbaselines\m@th
    \ialign{${\displaystyle ##}$\hfil&&\ ${\displaystyle ##}$\hfil\crcr
    \mathstrut\crcr\noalign{\kern-\baselineskip}
    \noalign{\smallskip}
    #1\crcr\mathstrut\crcr\noalign{\kern-\baselineskip}}}}

\newcommand\be{\begin{equation}}
\newcommand\ee{\end{equation}}

\def\f_eps{}
\def\f_ps{}
\def\jxlnull#1{}

\newcommand\figa{
  \begin{figure}
    \includegraphics[width=8.5cm]{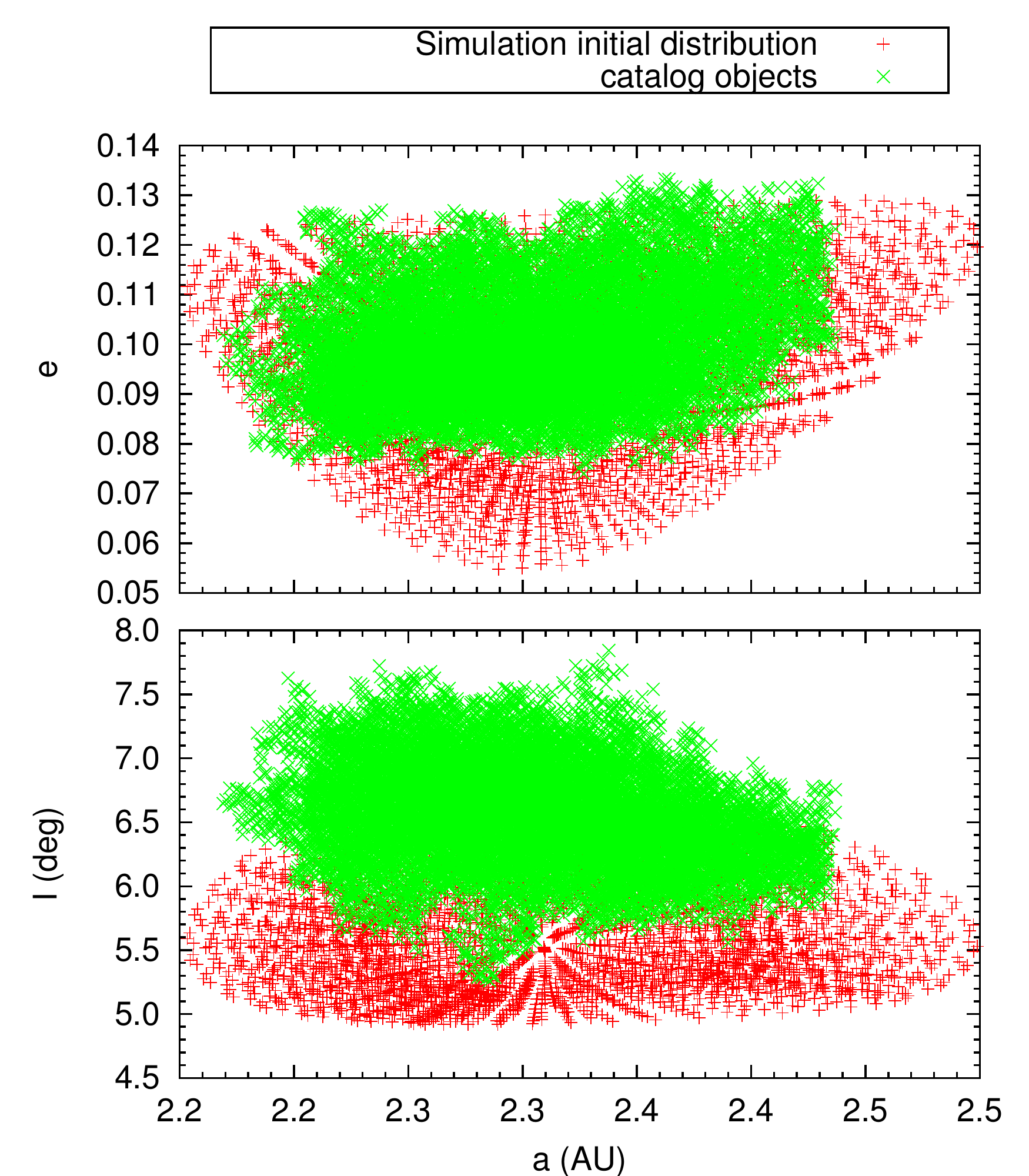} 
    \caption{Initial distribution of the synthetic Vesta family in the $(a,e)$ plane (top) and the $(a,I)$  plane (bottom).
      Real members proper elements \citep[from][]{2010PDSS..133N} are superimposed for comparison.}
    \llabel{Figa}
  \end{figure}
}

\newcommand\figb{
  \begin{figure}
    \includegraphics[width=8.5cm]{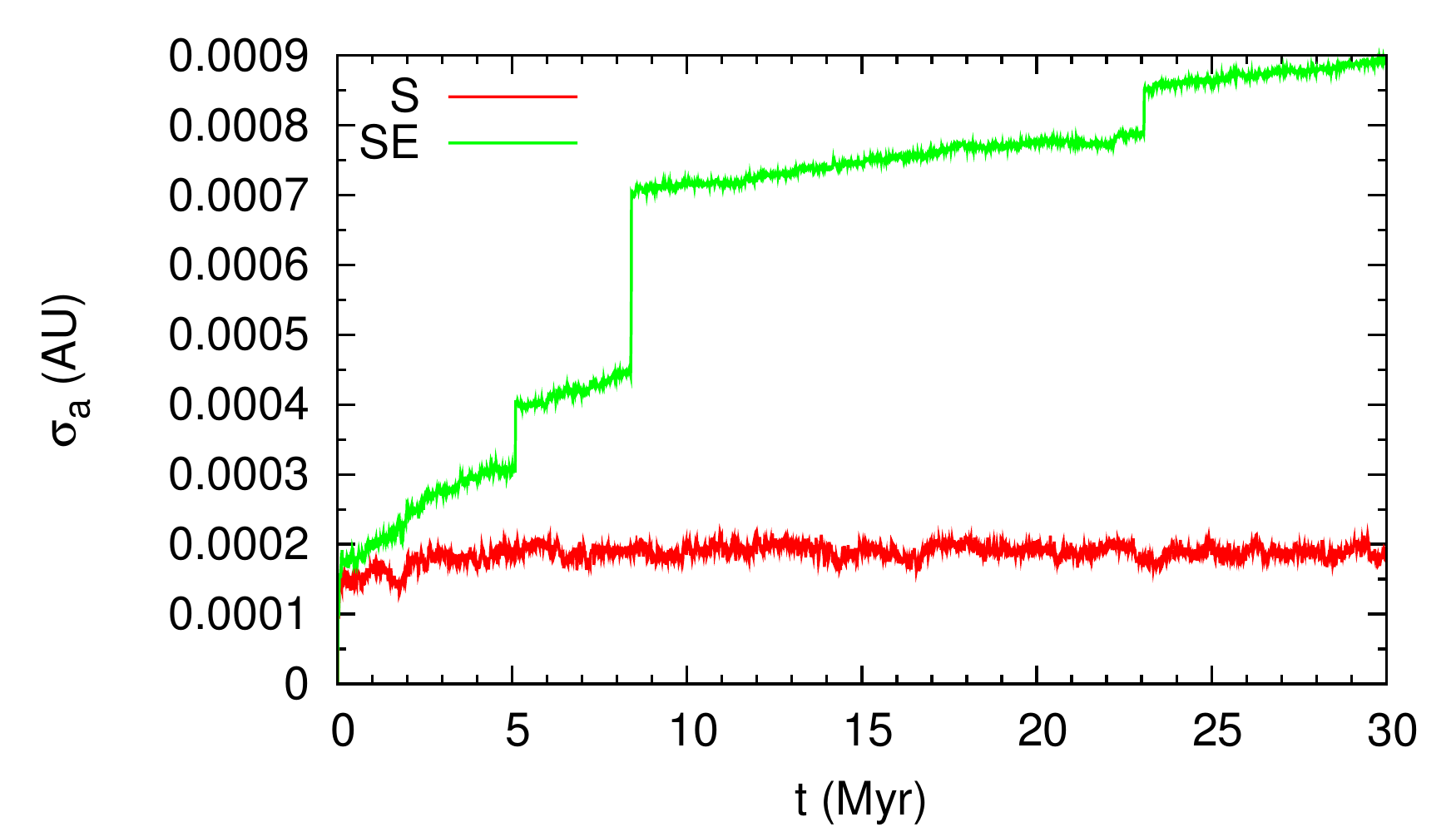}
    \caption{Evolution in time of the semi-major axis diffusion in $S$ and $SE$. We plot the standard deviation of the average of the minimum and the maximum values of a for each fragment over 10 kyr steps with respect to the initial values (see Eq.\ref{eq:sigma}) for asteroids whose semi-major axes are initially between 2.26 and 2.48 AU (in order to avoid strong resonances).}
    \llabel{Figb}
  \end{figure}
}

\newcommand\figc{
  \begin{figure}
    \includegraphics[width=8.5cm]{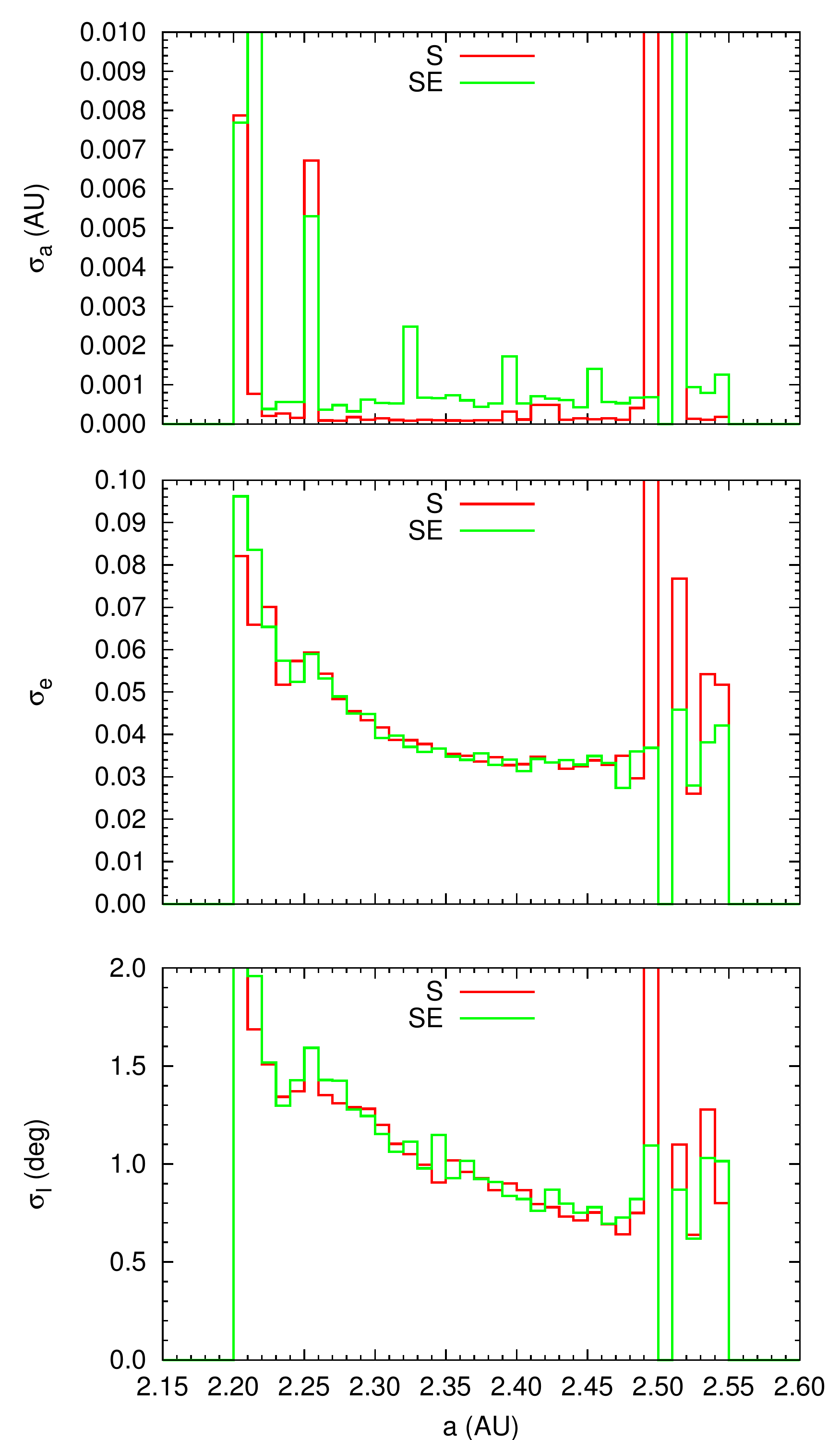}
    \caption{Semi-major axis dependency of the diffusion in semi-major axis (top), eccentricity (middle) and inclination (bottom) in $S$ and $SE$. For each band of 0.01 AU, we plot the standard deviation of the average of the minimum and the maximum values of $a$ (respectively $e$ and $I$) during the last 10 kyr of the simulations with respect to the initial values (see Eq.\ref{eq:sigma}).}
    \llabel{Figc}
  \end{figure}
}

\newcommand\figd{
  \begin{figure}
    \includegraphics[width=8.5cm]{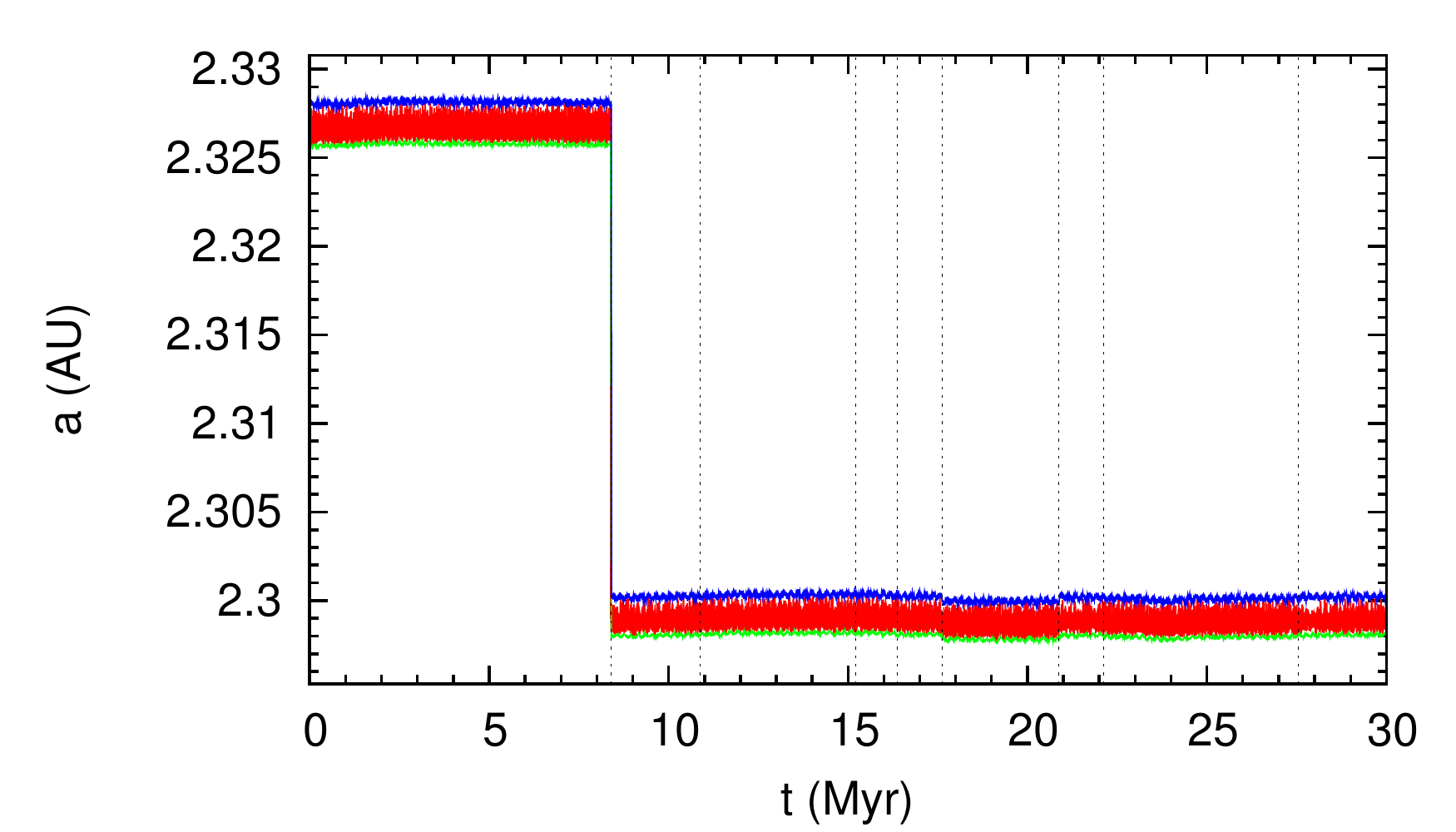}
    \caption{Semi-major axis evolution of a fragment which underwent a very close encounter (minimum distance of $3.4\times 10^{-5}$ AU) with Vesta (at $t \approx 8.39$ Myr). We plot the maximum, minimum and instantaneous values of the semi-major axis of the fragment. All times of encounters closer than $10^{-3}$ AU are highlighted with dotted vertical lines.}
    \llabel{Figd}
  \end{figure}
}

\newcommand\figdbis{
  \begin{figure}
    \includegraphics[width=8.5cm]{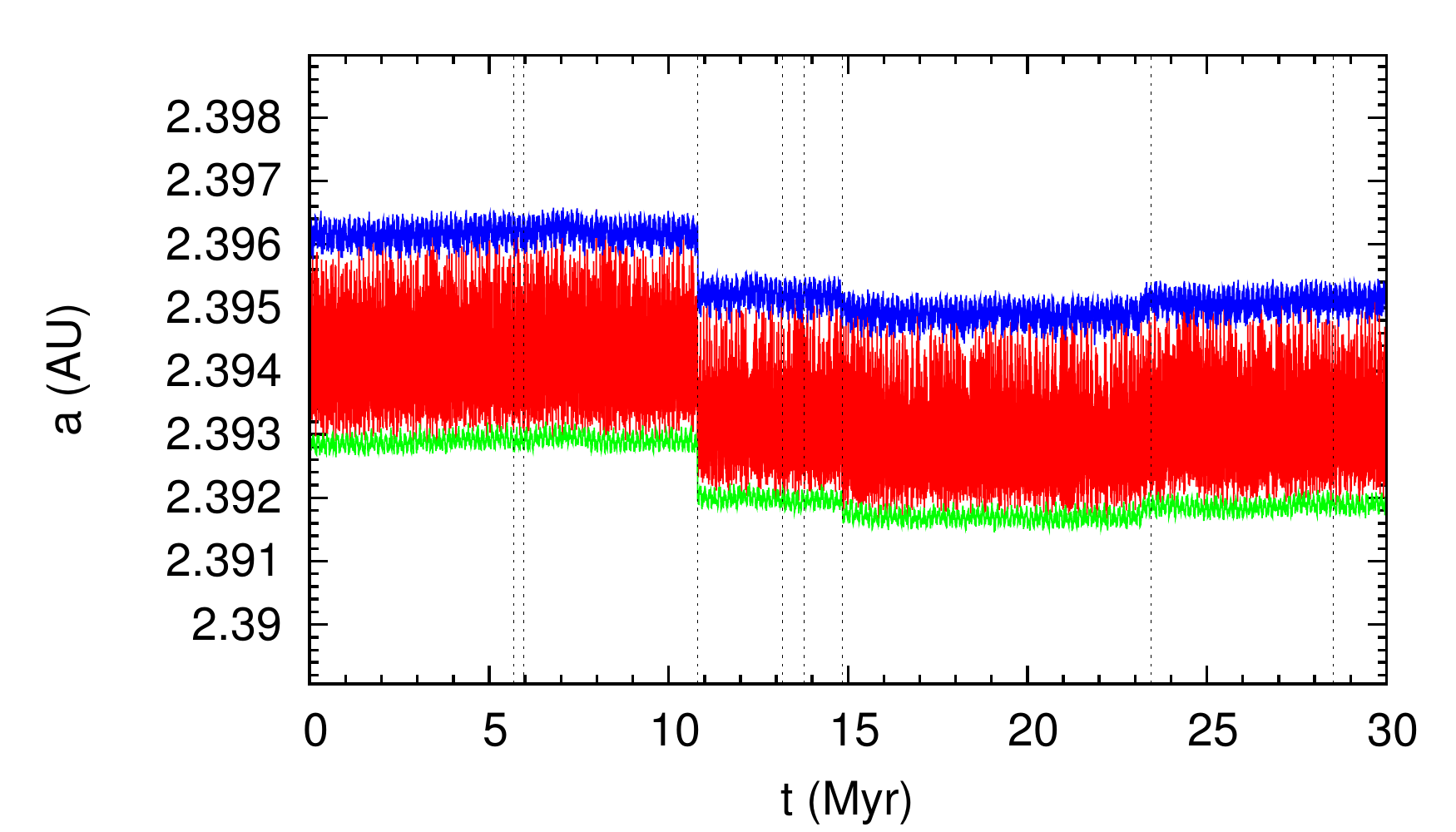}
    \caption{Semi-major axis evolution of a fragment which underwent various close encounters. We plot the maximum, minimum and instantaneous values of the semi-major axis of the fragment. All times of encounters closer than $10^{-3}$ AU are highlighted with dotted vertical lines.}
    \llabel{Figdbis}
  \end{figure}
}

\newcommand\fige{
  \begin{figure}
    \includegraphics[width=8.5cm]{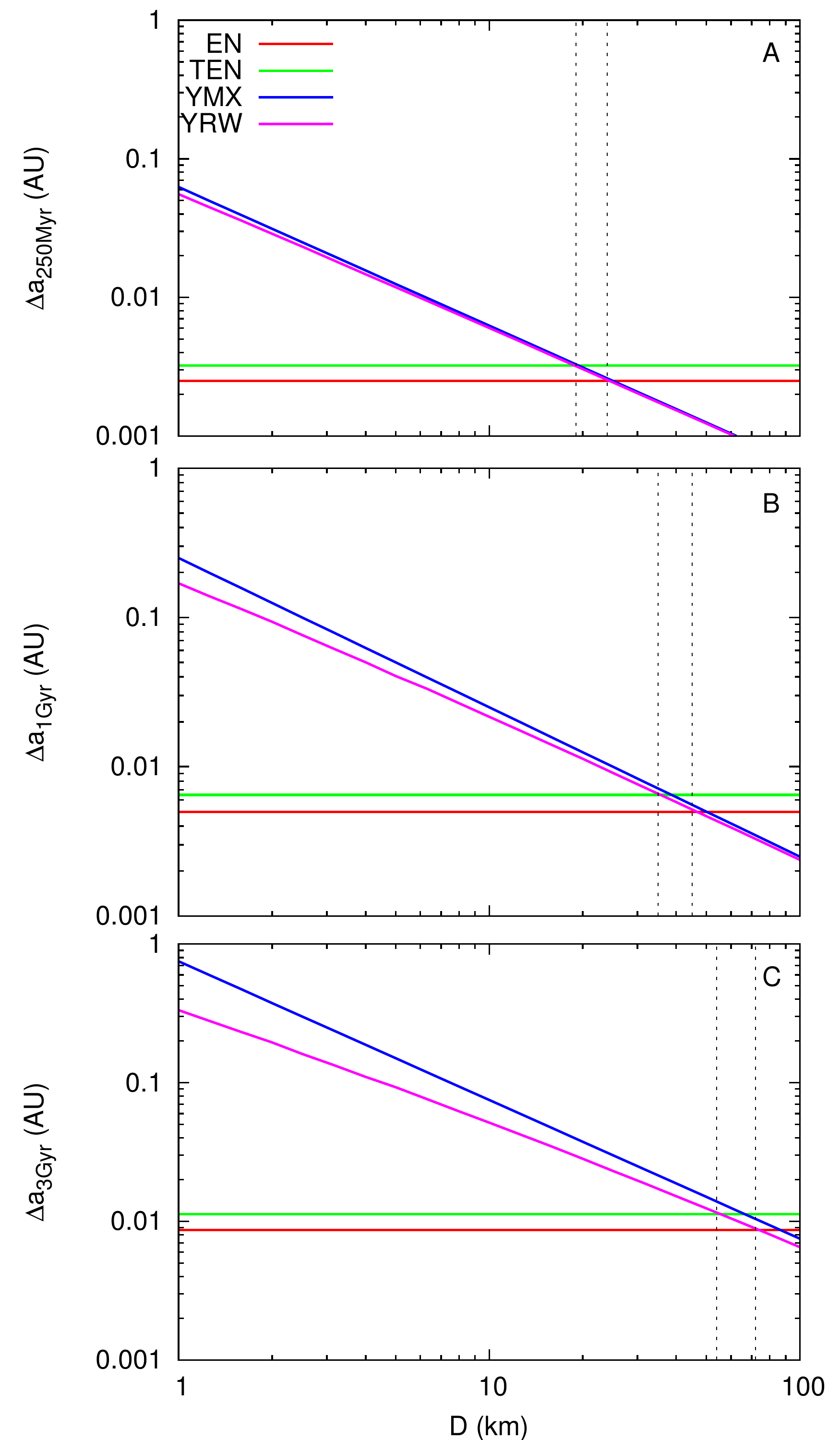}
    \caption{Comparison of the semi-major axis diffusion due to the close encounters and to the Yarkovsky effect after 250 Myr (A), 1 Gyr (B) and 3 Gyr (C).
      EN is the diffusion due to close encounters with the 11 asteroids considered in our simulations.
      TEN is the extrapolation we performed for the entire main belt ($\times 1.3$).
      YMX stands for the maximal diurnal Yarkovsky effect.
      YRW stands for the Yarkovsky effect with reorientations (random walk).}
    \llabel{Fige}
  \end{figure}
}

\newcommand\figf{
  \begin{figure}
    \includegraphics[width=8.5cm]{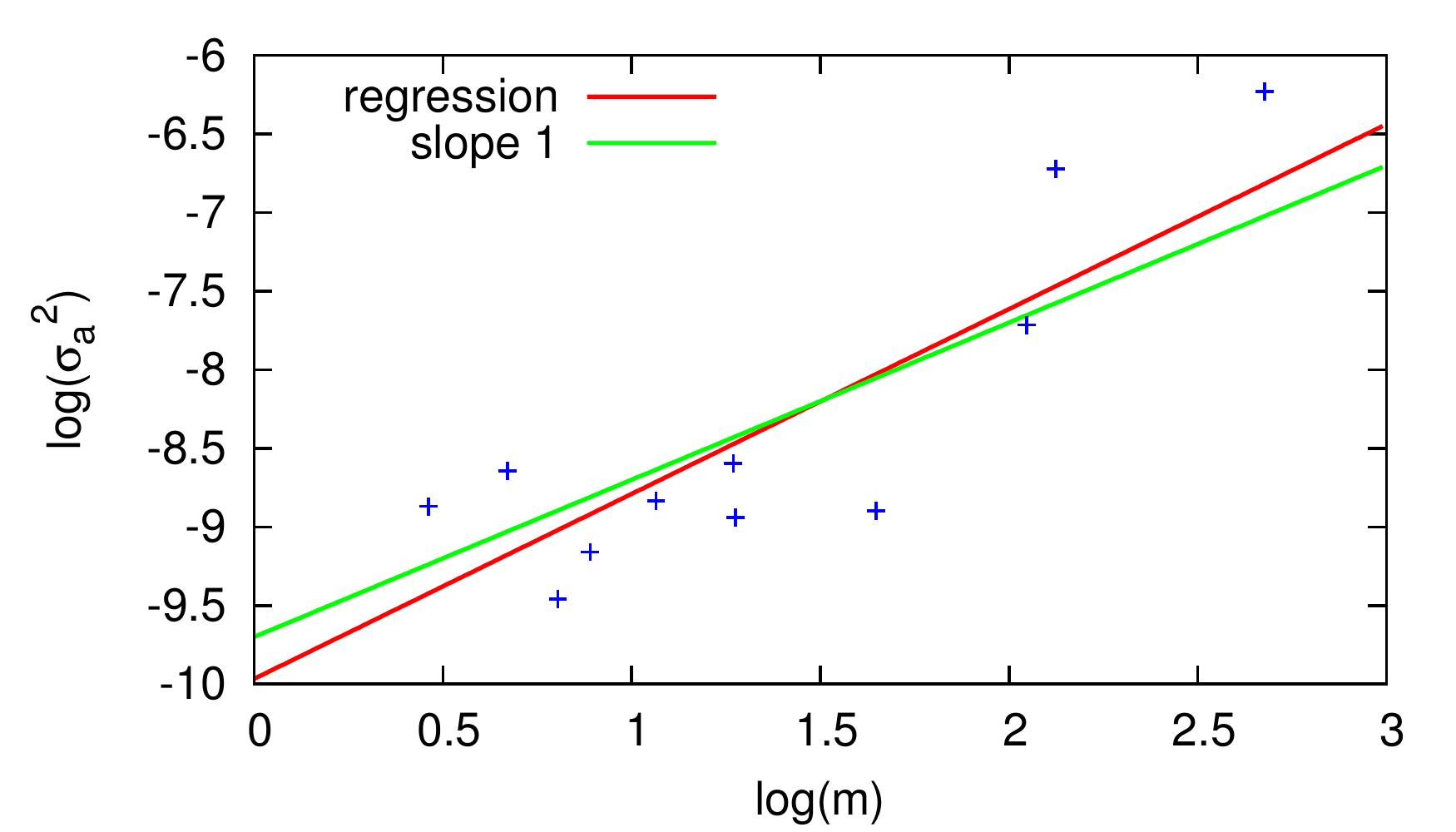}
    \caption{Contribution to the semi major axis diffusion due to each considered asteroid as a function of its mass.}
    \llabel{Figf}
  \end{figure}
}

\newcommand\figg{
  \begin{figure}
    \includegraphics[width=8.5cm]{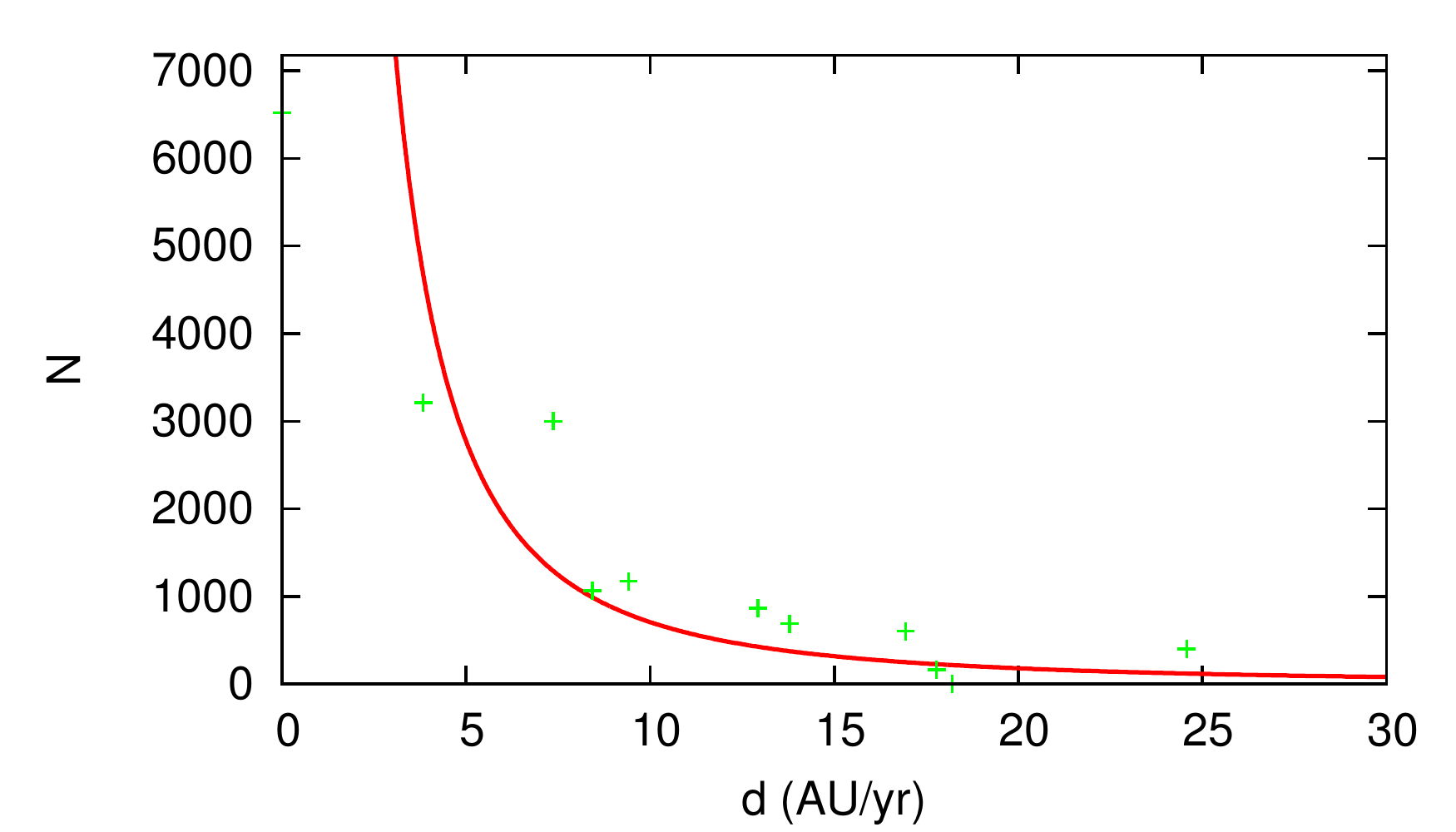}
    \caption{Number of close encounters ($< 10^{-3}$ AU) between Vesta family fragments and each of the 11 considered asteroids as a function of the distance in phase space between these asteroids and Vesta. We use the mean values of the semi-major axes, the eccentricities and the inclinations of the 11 asteroids during the 30 Myr of the simulation in order to compute the distances (see Eq.\ref{eq:dist}).}
    \llabel{Figg}
  \end{figure}
}

\newcommand\figh{
  \begin{figure}
    \includegraphics[width=8.5cm]{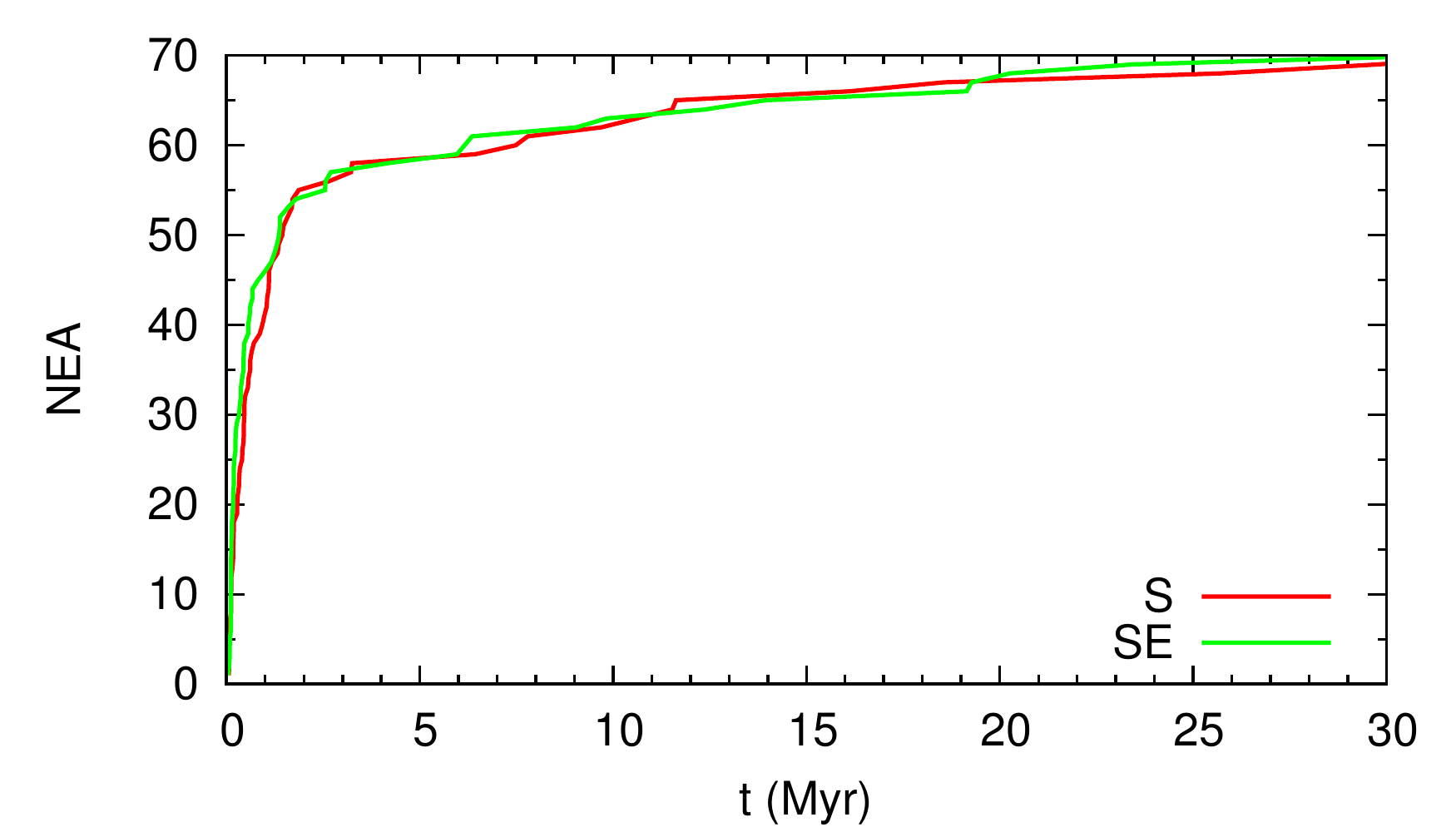}
    \caption{Number of near-Earth asteroids (NEA) among the synthetic Vesta family as a function of time in $S$ and $SE$. The criterion for considering a fragment as a NEA is based on the minimum value taken by the periastron. When this value is less than 1.3 AU the fragment is considered as a NEA.}
    \llabel{Figh}
  \end{figure}
}

\newcommand\taba{
  \begin{table} 
    \begin{center}
      \caption{Comparison of the contributions of the encounters with the 11 considered asteroids to the semi-major axis diffusion. All standard deviations are given in AU. The contributions are given as percentages of the total variance (which is additive unlike the standard deviation).}
      \begin{tabular}{rrrrrr}
\hline
Ast.   & $N_{events}$   & $\sigma_a[1]$ & $\bar\sigma_a[1]$ & $\bar\sigma_a(T_{sim})$ & Contrib.\\
 &   & $\times 10^5$ & $\times 10^5$ & $\times 10^5$ & (\%)\\
\hline
   1	 &  1175	 & 76.9 	 & 76.8	 & 51.4	 &    35.55 \\
   2	 &   401	 & 14.6 	 & 13.9	 &  5.4	 &     0.40 \\
   3	 &   868	 &  5.9 	 &  3.8	 &  2.2	 &     0.07 \\
   4	 &  6521	 & 43.8 	 & 43.6	 & 68.8	 &    63.63 \\
   7	 &  2999	 &  5.2 	 &  2.6	 &  2.8	 &     0.11 \\
  10	 &   162	 &  5.7 	 &  3.6	 &  0.9	 &     0.01 \\
  15	 &  1064	 &  5.6 	 &  3.4	 &  2.2	 &     0.06 \\
  19	 &  3210	 &  4.8 	 &  1.9	 &  2.1	 &     0.06 \\
 324	 &   603	 &  6.5 	 &  4.8	 &  2.3	 &     0.07 \\
 532	 &   688	 &  5.8 	 &  3.7	 &  1.9	 &     0.05 \\
 704	 &     4	 &  6.7 	 &  5.0	 &  0.2	 &     0.00 \\
\hline\hline
All	 & 17695	 & $-$	 & $-$	 & 86.3	 & 100 \\
\hline
      \end{tabular}
      \llabel{Taba}
    \end{center}
  \end{table}
}

\title{Chaotic diffusion of the Vesta family induced by close encounters with massive asteroids}
\author{J.-B. Delisle
\and J. Laskar
}

\institute{ ASD, IMCCE-CNRS UMR8028, Observatoire de Paris, UPMC, 
77 Av. Denfert-Rochereau, 75014 Paris, France
}
\offprints{J.-B. Delisle, \email{delisle@imcce.fr}}

\titlerunning{Chaotic diffusion induced by massive asteroids}
\authorrunning{Delisle \and Laskar}

\date{\today}
\abstract{
We numerically estimate the  semi-major axis  chaotic diffusion  of the  Vesta family asteroids induced by close encounters with 11 massive main belt asteroids :
(1) Ceres, (2) Pallas, (3) Juno, (4) Vesta, (7) Iris, (10) Hygiea, (15) Eunomia, (19) Fortuna, (324) Bamberga, (532) Herculina, (704) Interamnia.
We find that most of the diffusion is due to Ceres and Vesta.
By extrapolating our results, we are able to constrain the global effect of close encounters with all the main belt asteroids.
A comparison of this drift estimate with the one expected for the Yarkovsky effect shows that for asteroids whose diameter is larger than about 40 km, close encounters dominate the Yarkovsky effect.
Overall, our findings confirm the standard scenario for the history of the Vesta family.
}
  
\begin{document}

\maketitle

\section{Introduction}
It is now admitted that most of V-type near-Earth asteroids (NEAs) and howardite, eucrite and diogenite (HED) meteorites are fragments of a collision between Vesta and another object which also produced the Vesta family \citep[see][]{1997M&PS...32....3S}. V-type NEAs and HED meteorites are former members of the Vesta family that have been carried from the main asteroid belt to Earth-crossing orbits by the two main resonances in the neighborhood of Vesta : the $\nu_6$ and $3:1$ resonances with Jupiter.
The major issue in this scenario is the fact that the average life time of fragments in these two resonances is too short to explain the  mean age of HED meteorites and V-type NEAs \citep{1997M&PS...32..903M}.

The commonly accepted explanation for this is that NEAs spent most of their life time in the main asteroid belt before entering in resonance and being carried to their current orbits.
This supposes that a mechanism can induce a diffusion process that brings fragments to one (or both) of the two main resonances.
The Yarkovsky effect is now admitted to be the main mechanism explaining the diffusion of Vesta family members \citep[e.g.][]{2003Icar..162..308C}.

However other processes also contribute to the diffusion. It is in particular the case for the chaos induced by overlaps of mean motion resonances or close encounters between asteroids in the main belt. In the case of the resonances, \citet{1998M&PS...33..999M} showed that the chaotic zone around the $3:1$ resonance is too narrow to explain the needed diffusion.
Recently, \citet{2011A&A...532L...4L} showed that close encounters among massive asteroids induce strong chaos which appears to be the main limiting factor for planetary ephemeris on tens of Myr.
In the continuation of this work we concentrated our study on the diffusion induced by close encounters of Vesta family members with  massive asteroids.

This effect has already been studied for different asteroid families \citep[e.g. the Flora and Lixiaohua families][]{2002Icar..157..155N,2010CeMDA.107...35N}. The most exhaustive study (in terms of number of asteroids taken into account) concerned the Gefion and Adeona families \citep{2003Icar..162..308C}. The authors considered all 682 asteroids of radius larger than 50 km for encounters with members of the families. They concluded that the four largest asteroids ((1) Ceres, (2) Pallas, (4) Vesta and (10) Hygiea) had a much larger influence than all the 678 remaining asteroids so the latters can be considered as negligible.
The Vesta family has also been the object of such an analysis \citep{2007A&A...465..315C} but only close encounters with (4) Vesta were taken into account.
These different studies show that the effect of close encounters with massive asteroids depends a lot on the considered family and its position in phase space with respect to those of massive asteroids.

The purpose of this work is to explore further the effect of close encounters for the Vesta family by considering  the effect of  11 large  asteroids :
(1) Ceres, (2) Pallas, (3) Juno, (4) Vesta, (7) Iris, (10) Hygiea, (15) Eunomia, (19) Fortuna, (324) Bamberga, (532) Herculina, (704) Interamnia.
Moreover, we use the results from these interactions to extrapolate and obtain an estimate of the effect which would be raised by the whole main asteroid belt.
Finally we compare the semi-major axis diffusion induced by the Yarkovsky effect and the diffusion due to close encounters (in the case of the Vesta family).

\section{Numerical simulations}
We ran two sets of numerical simulations.
Both of them comprise the 8 planets of the Solar System, Pluto, the Moon and the 11 asteroids enumerated above.
In both simulations, we generate a synthetic Vesta family produced by an initial collision of another object with Vesta.
We do not model a realistic collision process but instead we consider a set of test particles initially at the same position as Vesta but with different relative velocities.
This is not a problem because we are mainly interested in this work in exploring the phase space in the region of the Vesta family.
The relative velocities of the fragments of collision are sampled both in norm and direction.
For the norm, we take 8 different values (every $50\textrm{ ms}^{-1}$) between the escape velocity at Vesta's surface (around $350\textrm{ ms}^{-1}$) and twice this value ($700\textrm{ ms}^{-1}$).
The inclination is sampled between $-50\deg$ and $50\deg$ every $10\deg$ (11 different values).
Finally, the direction in the invariant plane is taken every $10\deg$ (36 values).
This way we create 3168 test particles that represent the Vesta family.
We plotted the positions of these particles in the $(a,e)$ and $(a,I)$ planes (where $a$ is the semi-major axis, $e$ the eccentricity and $I$ the inclination) superimposed with the catalog published by \citet{2010PDSS..133N} of current Vesta family members (Fig.\ref{Figa}). 
Note that despite we did not compute the proper elements whereas the catalog uses them the two sets of points superimpose quite well.
Moreover we constructed an initial set of particles just after the original collision whereas the catalog exposes the remnants of the family after approximately 1.2 Gyr of evolution \citep[see][]{2007A&A...465..315C}.

\figa

In the first simulation (the reference simulation, $S$), the 11 asteroids are considered as test particles as well and thus do not perturb the fragments of collision.
In the second simulation ($SE$) the 11 asteroids are considered as the planets and thereby interact with members of the synthetic Vesta family.
Both solutions are integrated over 30 Myr using the symplectic integrator described in \citep{2011A&A...532L...4L} and references therein. The effects of all planets and Pluto are taken into account, 
as well as general relativity.

In order to be able to analyze the evolution of the test particles and in particular the impact of close encounters with the 11 asteroids we set up different logs.
We record for each particle its instantaneous orbital elements every 1 kyr.
We do not compute proper elements but instead we use the minimum and maximum values \citep[see][]{1994A&A...287L...9L} taken by $a$, $e$ and $I$ on time spans of 10 kyr.
Actually, we use the average of the minimum and maximum values as proper elements.
We also set up a log of close encounters for each particle. Each time that a particle passes within 0.01 AU from an asteroid, the asteroid number, the minimum distance of approach and the time of the encounter are recorded in the logs.

\section{Global overview of the diffusion}
The goal of this study is to understand whether the current flux of V-type asteroids coming from the main belt and carried to near-Earth orbits through strong planetary resonances can be explained by the chaos induced by close encounters in the main asteroid belt (or at least that these encounters contribute significantly).
The simplest way to check this assumption is to look in both $S$ and $SE$ simulations to the number of NEAs as a function of time and to see if the consideration of asteroidal interactions implies a greater number of NEAs.
We use the usual criterion to decide whether an asteroid is a NEA or not : if the periastron of its orbit become smaller than 1.3 AU the asteroid is considered as a NEA \citep[see for example][]{2002aste.conf..409M}.
In Fig.\ref{Figh} we plot the evolution of the number of NEAs in both simulations as a function of time.
We can see on this graph that the close encounters have not a major effect on the population of NEAs on the duration of the simulation.
\figh
However, we did not ran our simulations on the total age of the Vesta family (around 1.2 Gyr) and we can think that just after the collision that originated the Vesta family, the population of V-type NEAs is dominated by fragments of the collision that were directly injected in the two main resonances.
The diffusion process is supposed to provide these resonances on a much larger time scale.

Another way to look for the effect of close encounters is to compare the proper elements of the initial and final distribution of fragments in both simulations.
A usual way to do that is to compute the standard deviation of different proper elements (and in particular the semi-major axis) on the whole population of fragments at a time t with respect to the initial values of these elements \citep[see for example][]{2003Icar..162..308C} :
\be\llabel{eq:sigma}
  \sigma_x(t) = \sqrt{\frac{\sum_j (x_j(t) - x_j(0))^2}{N-1}},
\ee
where $x$ can be the semi-major axis $a$, the eccentricity $e$ or the inclination $I$ of Vesta family fragments.
As we mentioned it we use the averages of the minima and maxima as proper elements and in particular for the initial values we compute them on the first 10 kyr of the simulation. We do not use the initial conditions given in Fig.\ref{Figa} because they are only instantaneous values.

We measure the diffusion in semi-major axis only due to close encounters, thereby we do not  take into account resonances and in particular strong ones.
For this purpose we plotted the value of the standard deviation between the beginning and the end (after 30 Myr of evolution) for both simulations ($S$, $SE$) as a function of the initial (proper) semi-major axis. We divided the interval of semi major axis in bands of 0.01 AU and for each band we computed the standard deviation (of the semi-major axis, the eccentricity and the inclination) of the set of asteroids that are initially in this band.
The results of this calculation are given in Fig.\ref{Figc}. We can see that the strong resonance $3:1$ around 2.5 AU induces an important diffusion in both $S$ and $SE$ and that for semi-major axis lower than 2.25 AU both simulations are strongly affected by resonances \citep[see][for a list of resonances acting on the Vesta family]{2008Icar..193...85N}.
Note that in strong resonances most fragments are highly unstable and for a great part of them the calculation stops before the end of the simulation (collisions with a planet or the Sun or escapes from the Solar System). Thereby calculations of the diffusion between the beginning and the end of the simulation in such zones is not representative of their instability. For instance the diffusion is zero between 2.5 and 2.51 AU only because none of the fragments that were initially in this zone finished the simulation.
\figc

Between 2.26 and 2.48 AU, $S$ shows a very limited semi major-axis diffusion even if we can see small peaks (e.g. around 2.42 AU corresponding to the $1:2$ resonance with Mars).
$SE$ undergoes a more important diffusion in this band.
We  run our calculations on this band because it fulfills at the same time two important conditions : the diffusion due to resonances is limited and the band contains (at the beginning of the simulation) a sufficient amount of fragments (2617 of the 3168 fragments) which is important for the statistics.
Note that for the eccentricity and the inclination, the diffusion is hidden in the remaining oscillations due to the secular evolution of these elements induced by the planets and that we cannot see any difference between both simulations in Fig.\ref{Figc}.
However, this is not really a problem since in this work we are mainly interested in the semi-major axis diffusion.
\figb

The evolution (in time) of the standard deviation of the semi-major axis computed on the 2.26 - 2.48 AU band for both simulations is plotted in Fig.\ref{Figb}.
The standard deviation is approximately constant for $S$ which reinforces our choice of this band.
On the contrary, $SE$ clearly undergoes diffusion.
Analyzing Fig.\ref{Figb} gives us an approximation of the drift in semi-major axis due to asteroidal interactions during the simulation.
After 30 Myr of evolution, we have a standard deviation around $9\times 10^{-4}$ AU for $SE$ and $2\times 10^{-4}$ AU for $S$.
In order to evaluate the diffusion due to asteroidal interactions, we have to remove the diffusion obtained in the reference simulation (which is probably due to resonances and output sampling of the secular evolution of the semi-major axis) from the total standard deviation.
Note that the total variance (the squared of the standard deviation) is the sum of the variances of the different contributions.
Thus we obtain a standard deviation due to asteroidal interactions after 30 Myr of evolution of about $8.8\times 10^{-4}$ AU.
We also observe several jumps in the semi-major axis diffusion, which are due to very close encounters of a single fragment with one of the 11 considered asteroids.
Actually it has been shown \citep{2007A&A...465..315C} that the distribution of jump sizes is not a Gaussian but has thicker wings.
Thus the curve is not regular on the time scale of this simulation because of the presence of rare very important events.
We can find the same kind of effects on Fig.\ref{Figc}, the curve is not regular in the 2.26 - 2.48 AU band and shows peaks due to a few very close encounters.

\section{Diffusion induced by asteroids close encounters  }
The global analysis that we made does not allow us to separate the contributions of the 11 asteroids in the diffusion process and to extrapolate to the whole main asteroid belt.
In order to do that we have to examine logs of close encounters and estimate the diffusion resulting from each of these events and then make statistics.
All this analysis is still restricted to the 2.26 - 2.48 AU band in order to avoid the strong resonances as previously.

For each encounter recorded in the logs we compare the value of the semi-major axis before and after the encounter.
Fig.\ref{Figdbis} gives an example of the evolution of the semi-major axis of a fragment that underwent several close encounters which resulted in jumps of different sizes.
Fig.\ref{Figd} gives a more unusual example of the evolution of a fragment which underwent a very close encounter with Vesta resulting in a very important jump in semi-major axis.
\figdbis
\figd
Then for each of the 11 asteroids we calculate the standard deviation of the distribution of jump sizes induced by close encounters with this asteroid.
This gives a measure of the average diffusion resulting from a single encounter with the asteroid.
As we already noticed, it is more convenient to manipulate variances than standard deviations since the latter are not additive.
If we assume that close encounters are independent events, every fragment undergoes a random walk and after $N$ encounters, the total variance is multiplied by a factor $N$ :
\be\llabel{eq:sigN}
\sigma_a^2 [N] = N \sigma_a^2 [1].
\ee
This gives a measure of the average diffusion resulting from $N$ encounters with the considered asteroid.
We just have then to replace $N$ by the mean number of close encounters per fragment with the selected asteroid during the whole simulation and we obtain the diffusion (variance) due to this asteroid on the duration of the simulation.
The diffusion due to close encounters with all the 11 asteroids on the duration of the simulation is the sum of the variances of the 11 asteroids.
With the same reasoning as for Eq.\ref{eq:sigN} it is possible to extrapolate the value of the variance on longer time scales :
\be\llabel{eq:extrapt}
\sigma_a^2 (t) = \frac{t}{T_{sim}} \sigma_a^2 (T_{sim}).
\ee
Note that it as been shown \citep[e.g.][]{2002Icar..157..155N} that close encounters do not exactly result in a random walk (there are some correlations between successive encounters) and that the variance is not exactly linear with time (or number of encounters). It can, indeed, be written as a power law of the form :
\be
\sigma_a(t) = C t^B.
\ee
For a random walk : $B = 0.5$.
Numerical estimations of this coefficient $B$ for different families \citep[see][]{2002Icar..157..155N,2003Icar..162..308C,2010CeMDA.107...35N} show that it depends on the family considered and can even evolve with time.
However the values found in the literature range between 0.5 and 0.65 and we did not find any estimations in the case of the Vesta family so we decided to stick to the random walk hypothesis. 

Note that in this reasoning we do not take into account the influence of the noise in the calculation of jumps sizes.
Indeed, we observe on Fig.\ref{Figdbis} and Fig.\ref{Figd} that the minimum and maximum values of the semi-major axis are not constant outside the close encounters (jumps) but undergo oscillations which are due to remaining secular evolution.
It is possible to eliminate an important part of this noise by taking the minimum of minima and the maximum of maxima over greater time scales before and after the encounters.
In order to choose the best time interval to compute these extrema and to evaluate the remaining noise after this treatment we use the reference simulation $S$ and we calculate the standard deviation of the difference of semi-major axis before and after random times.
Since there are no close encounters thus no jumps in this simulation, this standard deviation measures only the remaining noise.
With this method we compute the standard deviation of the noise for different time intervals for the calculation of extrema and we find that the minimal value ($4.4\times 10^{-5}$ AU) is obtained for an interval length of 200 kyr.
This is the value we employ for all the following calculations.

When we compute the variance of jump sizes in $SE$ (for real close encounters) we are affected by the same noise.
The value we obtain for the variance is the sum of the variance of the real diffusion and the variance of the noise.
Thus the real diffusion resulting from close encounters is given by :
\be
\bar{\sigma}_a^2[1] = \sigma_a^2[1] - \sigma_{noise}^2.
\ee

Another problem we experienced is that the logs of close encounters record all encounters within 0.01 AU but it appears that at this distance the effect is too small and is completely hidden in the noise. Thus statistics are contaminated by false events (with no real jumps).
In order to avoid this contamination we  take into account only encounters within 0.001 AU which are more significant events.

Finally, when a fragment undergoes several close encounters spaced by less than 200 kyr, we are not able to separate the effect of each encounter.
We thus  ignore all such multiple encounters and we keep only single encounters in the calculation of the standard deviation of the size of the jumps.
However, when we compute the total diffusion on the duration of the simulation we use the total number of encounters that occur during the simulation as multiple encounters should not be ignored.

Table \ref{Taba} sums up the results we obtain for each asteroid and for the 11 asteroids together.
We give the numbers of encounter during the simulation (including multiple encounters), the standard deviation obtained from statistics on single encounters ($\sigma_a[1]$), the corrected values from the noise ($\bar\sigma_a[1]$) and the standard deviations on the duration of the simulation ($\bar\sigma_a(T_{sim})$). Finally, the percentages of contribution of the different asteroids are given in term of variance.
\taba
We obtain a total diffusion due to close encounters of $8.63\times 10^{-4}$ AU during the 30 Myr of the simulation (see table \ref{Taba}).
This number is to be compared with the drift rate obtained with the global approach ($8.8\times 10^{-4}$ AU).
The results given by both methods are in good agreement.

By using Eq.\ref{eq:extrapt} we can extrapolate our results to longer times :
\be
\bar\sigma_a(t) =  1.57\times 10^{-4} \sqrt{t\ \textrm{(Myr)}}\ \textrm{AU}.
\ee
Of course it is possible to do the same for the contribution of each asteroid.
This allows us to compare our results with the previous study of \citet{2007A&A...465..315C}.
In this article, the authors find a drift rate of $2.0^{+2.5}_{-2.0}\times 10^{-3}$ AU/(100 Myr) by considering only close encounters with Vesta.
If we extrapolate to 100 Myr and consider only Vesta we find a drift of $1.3\times 10^{-3}$ AU which is coherent with \citet{2007A&A...465..315C} findings.

If we look at the different contributions in table \ref{Taba} we see that close encounters with Ceres and Vesta together represent about 99\% of the total diffusion (in variance). Contributions from the 9 other asteroids seem negligible.
This is coherent with findings of previous studies on other asteroid families \citep[see in particular][]{2003Icar..162..308C}.
On the contrary, since Ceres represents 36\% of the diffusion this asteroid should not be neglected (like in \citet{2007A&A...465..315C}).
We believe that two main reason can explain these proportions of contribution : the mass of the asteroids and their proximity with the Vesta family members in phase space.
The mass must influence the importance of the effect of a single close encounter whereas the proximity in phase space influences the number of such events.
In order to check for these correlations we plotted the variance due to close encounters as a function of the mass of the considered asteroid (Fig.\ref{Figf}) and the number of events as a function of the distance in phase space of the considered asteroid with respect to Vesta (Fig.\ref{Figg}).
We use the definition of the distance in phase space introduced by \citet{1994AJ....107..772Z} :
\be\llabel{eq:dist}
  d = n a' \sqrt{\frac{5}{4}\left(\frac{\delta a}{a'}\right)^2 + 2 (\delta e)^2 + 2 (\delta sin i)^2}.
\ee
This distance has the dimensions of a velocity since it is an estimation of the ejection velocity that would be needed to carry two objects initially at the same position to their current positions.
In Eq.\ref{eq:dist}, we used the mean values of the semi-major axes, the eccentricities and the inclinations of the 11 asteroids during the whole simulation (30 Myr) in order to compute the distances.
We deduce from these plots that we indeed have an effect of the mass and the proximity in phase space but we cannot obtain a simple and precise law in order to evaluate the diffusion that would result from close encounters with other asteroids than the ones considered in our simulation.
\figf
\figg

However, we can constrain the global effect of the remaining objects of the main asteroid belt.
In our simulation we took into account the most massive objects of the main asteroid belt.
Actually, if we add the masses of the 11 asteroids of our simulation we obtain a total mass of $8.36\times 10^{-10} M_\sun$.  The estimated total mass contained in the main asteroid belt is about $15\times 10^{-10} M_\sun$ \citep[obtained from INPOP10a fits,][]{2011CeMDA.tmp..101F} while 
\citet{2002Icar..158...98K} gave $18\times 10^{-10} M_\sun$.
The most massive asteroid in the main belt is Ceres with $4.76\times 10^{-10} M_\sun$, thus the remaining mass (not considered in our simulation) represents about 1.5 to 2 times the mass of Ceres.
The total mass contained in the main asteroid belt is not well constrained but we can assume
the upper limit of this remaining mass to about twice the mass of Ceres.

In Fig.\ref{Figf}, the slope of the regression curve is about 1.2.
We plotted a straight line of slope 1 for comparison.
A slope of 1 means that two objects of masses 0.5 have the same effect than one of mass 1.
If the slope is higher than 1 it is more efficient to have a single object whereas if the slope is lower than 1 it is more efficient to have two objects.
Even if this slope is not well constrained in Fig.\ref{Figf} it is reasonable to consider that it is greater than 1 so it is more efficient to have one big object than to split it into smaller objects.
Thereby, we can assume that the total contribution of the remaining objects of the main belt will be less than twice the contribution of Ceres.

We still have the problem of the influence of the distance in phase space but it seems realistic to consider that the case of Vesta is singular since it is the parent body of the Vesta family and that the remaining objects must have a probability of close encounter closer to Ceres's one than Vesta's one (see Fig.\ref{Figg}).
Thus, the total variance due to encounters with all the asteroids of the main belt should not exceed (and is probably a lot smaller than) 1.7 times the value of the variance obtained in our simulation.
In terms of standard deviations it corresponds to a factor of about 1.3.

\section{Comparison with the Yarkovsky-YORP effect}
The Yarkovsky effect is a non-gravitational force which results from the interaction between asteroids and the solar radiation.
There are actually two versions of the Yarkovsky effect : a diurnal and a seasonal effect.
The diurnal effect can result in an increase or a decrease of the semi-major axis depending on the obliquity ($\epsilon$) of the asteroid ($\propto \cos \epsilon$), whereas the seasonal effect systematically decreases the semi-major axis ($\propto \sin^2 \epsilon$).
The diurnal effect is maximal when $\epsilon = 0\deg$ or $180\deg$ and is zero when $\epsilon = 90\deg$.
On the contrary, the seasonal effect is maximal when $\epsilon = 90\deg$ and zero when $\epsilon = 0\deg, 180\deg$.
Both effects are size dependent, for instance the diurnal effect is given by \citep[see in particular][]{2008Icar..193...85N} :
\be
\frac{da}{dt} = 2.5\times 10^{-4} \left(\frac{1\ \textrm{km}}{D}\right) \cos \epsilon \textrm{ AU/Myr}.
\ee

In addition to the Yarkovsky effect, we have to consider the YORP (Yarkovsky - O'Keefe - Radzievskii - Paddack) effect which also comes from the solar radiation but affects the rotational velocity and the obliquity of asteroids.
This effect depends a lot on the shape and the surface composition of the asteroid but it is possible to do some statistics to understand what is the most probable scenario \citep[see][]{2002Icar..159..449V,2004Icar..172..526C}. It has been shown \citep{2004Icar..172..526C} that for most of basaltic asteroids (more than $95\%$), the obliquity is asymptotically driven to $0\deg$ or $180\deg$ by the YORP effect.
The time scale for the YORP effect is about 10-50 Myr \citep[see][]{2002Icar..159..449V,2003Icar..163..120M}.
This means that on a 1 Gyr time scale (which is the order of magnitude for the age of the Vesta family \citep{2007A&A...465..315C}), we can consider that the obliquity is either $0\deg$ or $180\deg$ during the whole evolution and that the seasonal effect is negligible and the diurnal effect is maximal (in one or the other direction).

Nevertheless, the collisions must also be taken into account.
We have to distinguish whether a collision is destructive or not.
The collisional life time of a 1-10 km asteroid in the main belt is about 1 Gyr.
But non-destructive collisions are more frequent and can reorient the spin axis of the asteroid.
The characteristic time scale of such events is given by \citep[see][]{1998Icar..132..378F,2003Icar..163..120M} :
\be
\tau_{coll} = 15 \left( \frac{R}{1\ \textrm{m}} \right)^{1/2}\ \textrm{Myr} = 335 \left( \frac{D}{1\ \textrm{km}} \right)^{1/2}\ \textrm{Myr}.
\ee
For kilometer-sized asteroids, the time between two reorienting collisions is about 300 Myr, it means that there are a few (3 or 4) reorientations on the time scale of the age of the Vesta family. Thus it is likely that these reorientations will reduce the diffusion due to the Yarkovsky effect but not by a large factor.
Moreover, we can still consider that the YORP effect acts almost instantaneously between two reorientations and that the seasonal Yarkovsky effect is negligible and the diurnal effect is always maximal.

Taking into consideration these different results we estimate the diffusion of Vesta family members under the effect of Yarkovsky, YORP and collisions by a simple random walk process and the maximal diurnal Yarkovsky effect between each reorientations.
Fig.\ref{Fige} shows a comparison between the diffusion due to close encounters and the one due to the Yarkovsky effect as functions of the asteroid diameter at different times   (250 Myr, 1 Gyr and 3 Gyr).
For the close encounters, we plotted the diffusion obtained for the 11 asteroids and the extrapolation we deduced for the whole main asteroid belt.
For the Yarkovsky effect we plotted both the maximal diurnal effect (without any reorientation) and the result of the random walk process.
We can see that reorienting collisions do not affect much the magnitude of the Yarkovsky effect for the range of diameters we are interested in, even at 3 Gyr.
We recall that the estimated age of the Vesta family is about 1 Gyr \citep{2007A&A...465..315C}.
For this duration, close encounters are dominant for the diffusion of asteroids larger than $40\pm 5$ km (Fig. \ref{Fige}),  but below this value the Yarkovsky effect is prevailing. For a diameter of 1 km the Yarkovsky effect is about 25 times greater than the effect of close encounters.

Regarding the time evolution of these effects, we assume in this work that close encounters generate a random walk process that evolves as $\sqrt{t}$ whereas the maximal Yarkovsky effect is linear with time.
When we take into account reorientations for the Yarkovsky effect we also have a random walk process so the evolution law should be proportional to $\sqrt{t}$.
However this is an asymptotic law which is valid when a great number of random walk steps is reached.
In the case of reorienting collisions the characteristic time is about 300 Myr so there are only a few steps per Gyr and the asymptotic law is invalid on this time scale.
It means that for the time scale we are interested in, the Yarkovsky effect evolves faster than $\sqrt{t}$ so faster than close encounters.
This is exactly what we observe on Fig.\ref{Fige} : for 250 Myr the close encounters take a more important part in the diffusion process (equivalence of the two effects for D=19 km) whereas for 3 Gyr the Yarkovsky effect is even more prevailing (equivalence for D=54 km).
\fige

\section{Conclusion}
In this work we evaluate the effect of close encounters between Vesta family members and massive asteroids on the semi-major axis diffusion of the family.
We calculate this effect by two methods. We first look at the collective effect of  close encounters and deduce an estimation of the semi-major axis drift due to the 11 massive asteroids taken into account here.
Then we study individually the close encounters and make statistics over these events.
With this second method we are able to separate the contribution of each of the 11 asteroids to the diffusion.
We show that both methods give comparable results and that our findings are compatible with previous work on the Vesta family \citep{2007A&A...465..315C}.
We use the results we obtained for the 11 asteroids to extrapolate and constrain the diffusion that would result from close encounters with all the main belt asteroids.
We show that this diffusion should not exceed 1.3 times the diffusion due to the 11 asteroids.
Finally, we compare the diffusion due to close encounters with the Yarkovsky-driven drift of the semi-major axis of asteroids in the 1-100 km range (in term of asteroids diameters).
We find that both effects are equivalent over 1 Gyr for a diameter of   $40\pm 5$ km. For smaller asteroids the Yarkovsky effect dominates the semi-major axis diffusion and for larger asteroids close encounters become more important.
Thus we confirm that although asteroids close encounters have a significant influence, the main mechanism of semi-major axis diffusion that transports main belt asteroids (and especially V-type asteroids) to Earth-crossing orbits via strong resonances is the Yarkovsky effect.
\begin{acknowledgements}
This work was supported by ANR-ASTCM, INSU-CNRS, PNP-CNRS, and CS, Paris Observatory.
Part of the computations were made at CINES/GENCI.
\end{acknowledgements}

\bibliographystyle{aa}
\bibliography{biblio}
\end{document}